# Using X-rays to Determine Which Compact Groups Are Illusory

Jeremiah P. Ostriker, Lori M. Lubin

Princeton University Observatory, Peyton Hall, Princeton, NJ 08544-0001

Lars Hernquist[1]

Lick Observatory, U.C. Santa Cruz, Santa Cruz, CA 95064

astro-ph/9411052  14 Nov 94

---

[1]Sloan Foundation Fellow, Presidential Faculty Fellow

## ABSTRACT

If the large-scale galaxy distribution is filamentary, as suggested by some observations and recent hydrodynamical simulations, then lengthwise views of filaments will apparently produce compact groups (CGs) that are in reality stretched out along the line of sight. This possibility has been advocated recently by Hernquist, Katz & Weinberg (1994).

Here, we propose a test for this hypothesis using X-ray emission from CGs. The observable quantity $Q \equiv L_x a_p^3 / L_g^2 T_x^{1/2}$ should be proportional to the axis ratio of the group, $a/c$, where $a$ and $c$ are the long and short axis of a prolate distribution, $a_p$ is the radius of the group projected onto the sky, $L_x$ is the bolometric X-ray luminosity, $L_g$ is the group blue luminosity, and $T_x$ is the gas temperature. We find that the distribution of $Q$ is consistent with the notion that many spiral-rich CGs with unusually small values of $(a/c)$ are frauds, i.e. that the values of $Q$ are anomalously small. An alternative possibility is that CGs are gas-poor relative to rich clusters; however, this can be tested using the Sunyaev-Zeldovich effect. If the groups have a normal ratio of gas-to-total mass, but are simply stretched out along the line of sight, a Sunyaev-Zeldovich signal should be detectable.

## 1. Introduction

"Compact Groups" (CGs) of galaxies are typically defined as associations of at least four relatively large galaxies within a circular area, projected onto the sky, having a radius $\lesssim 100\ h^{-1}\ kpc$. Exactly 100 such systems have been identified by Hickson (1982, 1993) and his collaborators. The groups in his sample are generally referred to as Hickson Compact Groups (HCGs), and are numbered as in Hickson (1993). The physical nature of compact groups, their evolutionary status in the face of dynamical friction, their puzzlingly high $M_{baryon}/M_{total}$ ratios, and even the reality of these systems are uncertain in the extreme. Recently, Hernquist, Katz & Weinberg (1994; hereafter, HKW) suggested, on the basis of new hydrodynamic simulations, that a significant fraction of HCGs are not physically bound systems, but rather are chance projections of galaxies separated along the line of sight by distances far greater than the $\sim 39\ h^{-1}\ kpc$ projected pairwise separations between their members (e.g. Hickson et al. 1992). The extensive literature concerning the observations and physical modeling of these puzzling systems is carefully reviewed in the papers cited above and thus, for brevity, will not be discussed here. Our purpose is to suggest a new test, based on X-ray observations, to determine whether or not HKW's hypothesis is correct. On the basis of our analysis, the answer appears to be in the affirmative: most of the compact groups containing spirals are probably frauds.

It is worth noting a physical reason, pointed out by HKW and distinct from prior cited arguments, why CGs are suspicious. The redshift surveys performed by the CfA and Arecibo groups (e.g. Geller & Huchra 1989; Giovanelli & Haynes 1993) show that galaxies are not distributed randomly even on large scales, but are organized into walls and filaments. Recent hydrodynamical simulations of the formation of cosmologic structure (*e.g.* Katz, Hernquist & Weinberg 1992; Cen & Ostriker 1992, 1993) contain such filaments. These studies, as well as the recent work by Gnedin (1994) and Bertschinger & Jain (1994), indicate that collapsing and intersecting Zeldovich pancakes are optimal sites for galaxy formation, providing a physical mechanism to explain the striking results provided by galaxy surveys. It is clear on statistical grounds that spurious high surface density projected groups will be more likely if galaxies are formed in chains or filaments than if they are distributed randomly or in spherically symmetric ensembles. HKW showed that chance, edge–on sightings of long galaxy chains are, in fact, roughly consistent with the observed frequency of HCGs, if the numerical modeling of HKW is, in this respect, a reliable caricature of the real Universe.

Here, we propose a new observational test of the physical nature of compact groups. Suppose that a CG is contained within a prolate ellipsoid of semi-minor axis $a$ and semi-major axis $c$ and is viewed along its major axis. As an example, assume that the gas in the CG is distributed according to the density profile

$$\rho = \rho_o(1 + x^2/a^2 + y^2/a^2 + z^2/c^2)^{-1}, \tag{1}$$

where the exact form is taken only for the sake of definiteness. It will be seen below that our arguments are dependent only on the shape factor $(c/a)$ and not on the detailed form of the gas distribution. The X–ray luminosity $(L_x)$ of the gas is proportional to $f(T)\,\rho^2$ Volume $\propto f(T)\,M_{gas}^2/$Volume, where $f(T)$ depends on the X-ray band observed, the X-ray spectrum, metallicity, and so forth, but is not a strong function of temperature and typically varies between the limits $T^{-1/2}$ to $T^{+1/2}$. Thus, for the assumed gas distribution, it follows that $L_x \propto f(T)\,M_{gas}^2\,a^{-3}\,(a/c)$. Hence, if all groups and clusters were spherical and had known radii, gas mass, and temperature, we would expect the quantity $Q' \equiv L_x\,a^3/[M_{gas}^2\,f(T)]$ to be an observed constant, and a variation of $Q'$ among clusters would indicate the distribution of $(a/c)$, all other factors being equal. Unusually low observed values of $Q'$ for certain systems would imply unusually small values of $(a/c)$; i.e. very elongated groups and clusters. Given our ignorance of some of the parameters entering into the definition of $Q'$, we define instead an observationally measurable quantity

$$Q \equiv \frac{L_x \, a_p^3}{L_g^2 \, f(T)} \propto \left(\frac{M_{total}}{L_g}\right)^2 \left(\frac{M_{gas}}{M_{total}}\right)^2 \left(\frac{a}{c}\right), \qquad (2)$$

where $a_p$ is the projected radius, $L_x$ is the bolometric X-ray luminosity, $L_g$ is the blue galactic luminosity, and $f(T)$ gives the temperature dependence for Bremsstrahlung (or Raymond-Smith) emissivity. We adopt a Hubble constant of 100 $h \, km \, s^{-1} \, Mpc^{-1}$ and note that $Q$ scales as $h^{-1}$. It has been known for some time that the scaling relation $L_x \propto L_g^2$ is approximately valid for groups and clusters (Jones & Forman 1992). The physical origin of this relation is that $a_p$ and $f(T)$ range far less for real physical systems than do $L_x$ and $L_g$, so the near constancy of $Q$ for normal X-ray clusters implies this scaling between $L_x$ and $L_g$.

Now that we have defined an observable quantity $Q$, we can examine its distribution among known groups and clusters of galaxies. Neglecting for the moment correlations amongst observables, the variance in Log $Q$ should be composed of twice the variances in Log ($M_{total}/L_g$) and Log ($M_{gas}/M_{total}$), if clusters and groups are roughly spherical, with an extra dispersion produced by elongated systems.

## 2. Results

In this study, we use all rich clusters (12 in total) which have accurately measured total blue luminosity (Oemler 1974; Kent & Gunn 1982; Tully & Shaya 1984; Postman et al. 1986, 1988; West et al. 1989; Hughes 1989; Bird et al. 1993), and X-ray temperature and bolometric X-ray luminosity (Edge et al. 1990; Edge & Stewart 1991; Henry & Arnaud 1991; David et al. 1993). We also include two Morgan poor clusters, MKW4 and AWM4 (Morgan et al. 1975; Albert et al. 1977; Kriss et al. 1983). For the projected radial extent ($a_p$), we take it to be the radius which contains 50% of the total X-ray emission, and is approximately 1.0 $h^{-1} \, Mpc$ for rich clusters and 0.5 $h^{-1} \, Mpc$ for poor clusters.

We compare these clusters with the optical and X-ray properties of 10 Hickson CGs (Hickson 1982; Hickson et al. 1992) and the CfA group NGC 2300 (Huchra & Geller 1982). The X-ray data for these groups is taken from the ROSAT observations of Pildis, Bregman & Evrard (1994; hereafter, PBE). In only two of their CGs and the NGC 2300 group was diffuse X-ray emission detected. Accurate estimates of the diffuse X-ray emission from the other 8 CGs could not be obtained, either because there was no measurable diffuse emission or because the diffuse emission could not be separated from that from individual galaxies (PBE). For these 8 groups, we calculate *upper limits* to the diffuse component by summing the X-ray luminosities of all the member galaxies. When required, a typical

X–ray temperature of 1 $keV$ is assumed. In the cases where an X–ray surface brightness profile for the group can be estimated, we take $a_p$ to be the radius which contains 50% of the X–ray emission. (Note that $a_p = a\sqrt{3}$ for the density distribution given in Eq. 1.) Otherwise, $a_p$ is taken to be the optical extent of the CGs, which is roughly coincident with the other definition when both measures are available. Relevant parameters of these groups, including spiral fraction, total blue luminosity, projected radial extent, and X-ray luminosity are listed in Table 1. We also indicate which groups have only upper limits to their diffuse X-ray emission.

In Figure 1a, we compare the X-ray and optical properties of the poor and rich clusters with the CGs. For each system, we show the bolometric X-ray luminosity times the projected radial extent cubed as a function of total blue luminosity times $f(T)^{1/2}$. Here, $f(T)$ is taken to be $f(T) = (kT)^{1/2}$. The filled black circles indicate poor and rich clusters. The solid line shows a best-fit to the cluster data of the form Log $(L_x a_p^3) = 2 \times$ Log $(L_g f(T)^{1/2}) - 24.82$. The open circles are group data from PBE, where open circles with arrows pointing downwards indicate upper limits to the diffuse X–ray emission. We note that previous analyses of ROSAT observations of HCG 62 (Ponman & Bertram 1993) and ROSAT and ASCA observations of the NGC 2300 group (Mulchaey et al. 1993; Mushotzky 1994) yield values of $a_p$ and $L_x$ discrepant from those of PBE, which is likely due to different background subtractions and assumptions about metallicity (see PBE). These earlier values are $a_p = 0.09\ h^{-1}\ Mpc$ and $L_x = 0.0275 \times 10^{44}\ h^{-2}\ ergs\ s^{-1}$ for HCG 62 and $a_p = 0.03\ h^{-1}\ Mpc$ and $L_x = 0.0058 \times 10^{44}\ h^{-2}\ ergs\ s^{-1}$ for NGC 2300, where $a_p$ is again approximated as the radius containing 50% of the X-ray emission. These measurements are indicated in Figure 1a by filled triangles.

In Figure 1b, we show a histogram of the values of Log $Q$ (Eq. 2), measured in $10^{44}\ ergs\ s^{-1}\ Mpc^3\ L_\odot^{-2}\ keV^{-1/2}$. The poor and rich clusters are denoted by the open histogram with the solid line indicating the best-fit gaussian which has a mean $-24.82$ and a standard deviation 0.41 (indicated by the bar). The histogram shown as horizontal lines denotes the groups with detected diffuse X–ray emission (PBE), while the histogram shown as vertical lines (with arrows) represents those groups with upper limits. The downward arrow near the top of the figure shows where a deviation of $3\sigma$ from the mean of the clusters would lie. Only one compact group, HCG 62 (which contains no spirals), is marginally consistent with the clusters. Values of Log $Q$ and the number of $\sigma$ deviations of each group from the mean are indicated in Table 1.

## 3. Conclusions

For normal rich and poor clusters, the quantity $Q$ defined by Eq. (2) has, in fact, a rather small variance; the dispersion in Log $Q$ is only $\sigma = 0.41$. If we include with the normal groups the CGs containing only elliptical galaxies [on the grounds that ellipticals are typically found in truly dense regions (Dressler 1980)], then the dispersion increases to only 0.52. The remaining CGs, each of which contains at least one spiral, are all *more* than $3\sigma$ away from the peak, with a median separation from the normal groups of more than $8\sigma$. This provides *prima facie* evidence that most of the CGs containing spirals are in fact low density ensembles stretched out along the line of sight.

Of course, other interpretations are possible. If $Q$ is small for some group of objects, it may be the case, as can be seen from Eq. (2), that $(a/c)$ is of order unity but that either $(M_{total}/L_g)$ or $(M_{gas}/M_{total})$ is systematically lower for this subset. The first possibility can be excluded on the grounds that the values of $(M_{total}/L_g)$ determined for many of these groups, on the assumption that $(a/c) \approx 1$, are large, not small, with typical values $M_{total}/L_g \approx 100 - 500 \ h^{-1} \ M_\odot/L_\odot$ (Hickson et al. 1992; Mulchaey et al. 1993; Ponman & Bertram 1993; PBE).

But, could it be that $(M_{gas}/M_{total})$ is much lower for CGs than it is for clusters? In systems with a small escape velocity, explosive events (e.g. multiple supernovae) should be able to eject gas more easily than in the great clusters (Yahil & Ostriker 1973, Wyse & Silk 1985), so such an effect is not physically implausible. But a straightforward test is available to check this option. The Sunyaev–Zeldovich effect is proportional to the integral of the pressure along a line of sight through a cluster. Specifically,

$$\begin{aligned}
\frac{\delta T}{T} &= \frac{-2\sigma_T}{m_e c^2} \int_0^s kT n_e dS \\
&\approx \frac{-2\sigma_T k <T>}{m_p m_e c^2} \frac{0.4 M_{gas}}{\pi a_p^2} \\
&\approx 1.2 \times 10^{-5} \left[ \frac{(\frac{M_{gas}}{M_{total}})}{0.07 \ h^{-1.5}} \right] \left[ \frac{(\frac{M_{total}}{L_g})}{200 \ h^{-1} \ M_\odot/L_\odot} \right] \left[ \frac{(\frac{L_g}{a_p^2})}{2 \times 10^{13} \ L_\odot/Mpc^2} \right],
\end{aligned}$$

where we assume a characteristic group temperature of 1 $keV$ and that $\sim 40\%$ of the total $M_{gas}$ lies within a projected radius $a_p$.

Thus, for the values of $(L_g/a_p^2)$ listed in Table 1, a detectable Sunyaev-Zeldovich effect is expected on the assumption that $(M_{gas}/M_{total})$ and $(M_{gal}/L_g)$ are in the normally observed ranges.

The data set we have used in our preliminary investigation is small, and significant observational uncertainties remain. In addition, as mentioned above, there is still the possibility that $(M_{gas}/M_{total})$ is low for CGs. However, the distribution of $Q$ values is so suggestive and the segregation of CGs with spirals to very low values of $Q$ so stark that we believe further work to confirm or reject the analysis presented here is warranted.

We thank Neta A. Bahcall, Renyue Cen, and Neal Katz for useful discussions. We also thank Joel Bregman, John Mulchaey, and Rachel A. Pildis for providing early access to group X-ray data. LML thanks Richard Mushotzky for providing bolometric corrections for the ROSAT data and preliminary ASCA results and Andrew Lange and Anthony Readhead for providing recent experimental results regarding the Sunyaev–Zeldovich effect. LH received support from the Alfred P. Sloan Foundation, NASA Theory Grant NAGW–2422, and from the NSF under Grants AST 90–18526, ASC 93–18185, and the Presidential Faculty Fellows Program. LML acknowledges support from NASA through Grant NGT–51295. JPO acknowledges support from NASA Grant NAGW–2448 and NSF Grant AST91–08103.

# REFERENCES

Albert, C.E., White, R.A., Morgan, W.W. 1977, ApJ, 211, 309

Bertschinger, E.W. & Jain, B. 1994, ApJ, 431, 486

Bird, C.M., Dickey, J.M., & Salpeter, E.E. 1993, ApJ, 404, 81

Cen, R., & Ostriker, J.P. 1992, ApJ, 399, L113

Cen, R., & Ostriker, J.P. 1993, ApJ, 417, 415

David, L.P., Slyz, A., Jones, C., Forman, W., & Vrtilek, S.D. 1993, ApJ, 412, 479

Dressler, A. 1980, ApJ, 236, 351

Edge, A.C., & Stewart, G.C. 1991, MNRAS, 252, 428

Edge, A.C., Stewart, G.C., Fabian, A.C., & Arnaud, K.A. 1990, MNRAS, 245, 559

Geller, M.J. & Huchra, J.P. 1989, Science, 246, 897

Giovanelli, R. & Haynes, M.P. 1993, AJ, 105, 1271

Gnedin, N.Y. 1994, ApJ, in press

Henry, J.P., & Arnaud, K.A. 1991, ApJ, 374, 410

Hernquist, L., Katz, N.S. & Weinberg, D.H. 1994 ApJ, in press (HKW)

Hickson, P. 1982, ApJ, 255, 382

Hickson, P. 1993, Ap. Lett. Comm. 29 1

Hickson, P., Mendes de Oliveira, C., Huchra, J.P. & Palumbo, G.G.C. 1992, ApJ, 399, 353

Hughes, J.P. 1989, ApJ, 337, 21

Huchra, J.P. & Geller, M.J. 1982, ApJ, 257, 423

Jones, C., & Forman, W. 1992, in : Clusters and Superclusters of Galaxies, ed. A.C. Fabian, NATO ASI Series (Dordrecht : Kluwer)

Katz, N.S., Hernquist, L. & Weinberg, D.H. 1992, ApJ, 399, L109

Kent, S.M., & Gunn, J.E. 1982, AJ, 87, 945

Kriss, G.A., Cioffi, D.F., & Canizares, C.R. 1983, ApJ, 272, 439

Morgan, W.W., Kayser, S., & White, R.A. 1975, ApJ, 199, 545

Mulchaey, J.S., David, D.S., Mushotzky, R.F., & Burstein, D. 1993, ApJ, 404, L9

Mushotzky, R. 1994, private communication

Oemler, A. 1974, ApJ, 194, 1

Pildis, R.A., Bregman, J.N. & Evrard, A.E. 1994, ApJ, submitted (PBE)

Ponman, T.J. & Bertram, D. 1993, *Nature*, 363, 51

Postman, M., Geller, M.J., & Huchra, J.P. 1986, AJ, 91, 1267

Postman, M., Geller, M.J., & Huchra, J.P. 1988, AJ, 95, 267

Tully, R.B., & Shaya, E.J. 1984, ApJ, 281, 31

West, M.J., Oemler, A., & Dekel, A. 1989, ApJ, 346, 539

Wyse, R.F.G. & Silk, J. 1985, ApJ, 296, L1

Yahil, A. & Ostriker, J.P. 1973, ApJ, 185, 787

## Figure Captions

Figure 1. Comparison of X-ray and optical properties of poor and rich clusters with small and compact groups. Panel (a) shows $L_x a_p^3$ versus $L_g f(T)^{1/2}$. The filled black circles indicate poor and rich clusters. The solid line gives the best-fit to the cluster data (see Sect. 2). The open circles are derived from the CG group data of PBE, where open circles with arrows pointing downwards indicate upper limits to the diffuse X-ray emission. The filled triangles indicate previous X-ray analyses of HCG 62 (Ponman & Bertram 1993) and the CfA group NGC 2300 (Mulchaey et al. 1993). In this panel, $L_x$ is measured in $10^{44}$ ergs/sec, $a_p$ is measured in Mpc, $L_g$ is measured in $L_\odot$, and $kT$ is measured in keV.

Panel (b) shows histograms of Log $Q$ (Eq. 2). The poor and rich clusters are indicated by the open histogram with the solid line providing the best-fit gaussian which has a mean $-24.82$ and a standard deviation $0.41$ (indicated by the bar). The horizontal-line histogram represents the PBE detections of diffuse emission from CGs (and the NGC 2300 group), while the vertical-line histogram (with arrows) represents upper limits on diffuse emission in the remaining CGs. Regions of histogram that have both horizontal and vertical lines imply that there are two separate bins; one for CGs detected and the other for CGs with an upper limit to the diffuse X-ray emission. The downward arrow near the top shows where a deviation of $3\sigma$ from the mean of the clusters would lie. Only one group, HCG 62 (all ellipticals), is marginally consistent with the clusters.

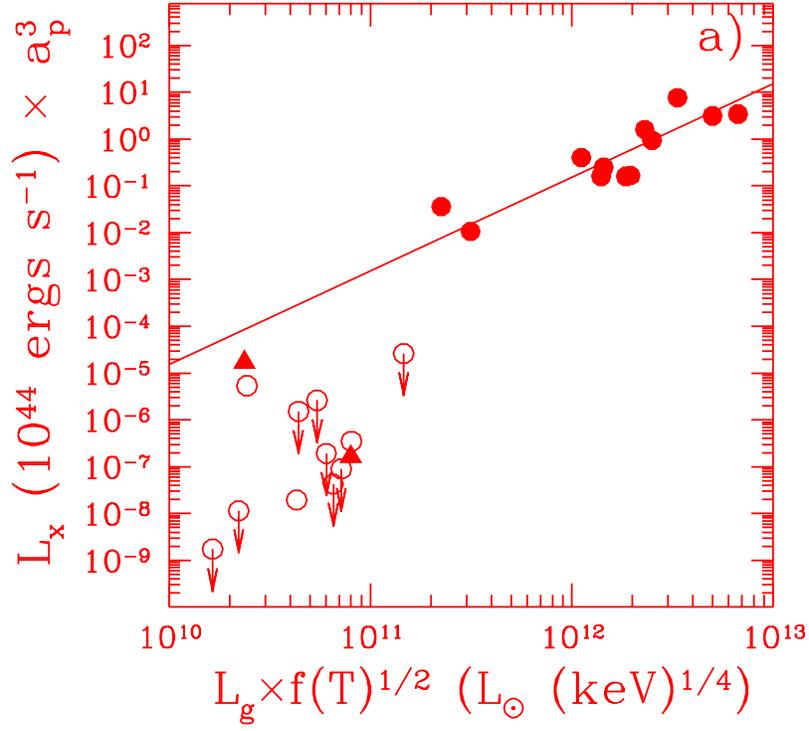

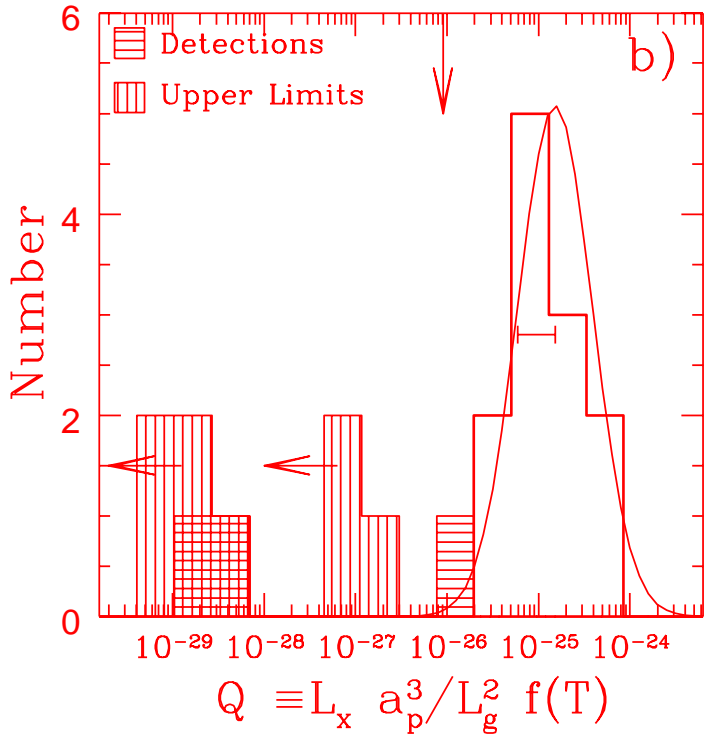

Table 1 : X-ray and Optical Properties of Compact Groups

| Group Name | # of spiral/total | $L_g$ (ergs s$^{-1}$) | $a_p$ (Mpc) | $L_x$ ($10^{44}$ ergs s$^{-1}$) | Log Q | $\sigma$ | Upper Limit |
|---|---|---|---|---|---|---|---|
| HCG 2 | 3/3 | $2.21 \times 10^{10}$ | 0.045 | 0.00013 | −28.63 | 9.3 | Y |
| HCG 4 | 1/3 | $5.43 \times 10^{10}$ | 0.044 | 0.03102 | −27.05 | 5.5 | Y |
| HCG 10 | 3/4 | $7.16 \times 10^{10}$ | 0.077 | 0.00020 | −28.76 | 9.6 | Y |
| HCG 12 | 1/5 | $6.54 \times 10^{10}$ | 0.044 | 0.00228 | −28.27 | 8.4 | Y |
| HCG 44 | 3/4 | $1.64 \times 10^{10}$ | 0.033 | 0.00005 | −29.19 | 10.7 | Y |
| HCG 62* | 0/3 | $2.43 \times 10^{10}$ | 0.100 | 0.00537 | −26.04 | 3.0 | N |
| HCG 68 | 1/5 | $4.31 \times 10^{10}$ | 0.039 | 0.00033 | −28.97 | 10.2 | N |
| HCG 93 | 3/4 | $6.53 \times 10^{10}$ | 0.066 | 0.00015 | −29.00 | 10.2 | Y |
| HCG 94 | 1/7 | $1.46 \times 10^{11}$ | 0.051 | 0.19800 | −26.91 | 5.1 | Y |
| HCG 97 | 2/5 | $4.41 \times 10^{10}$ | 0.075 | 0.00363 | −27.19 | 5.6 | Y |
| NGC 2300* | 2/3 | $8.20 \times 10^{10}$ | 0.076 | 0.00080 | −28.27 | 8.4 | N |

* Previous analysis of the X-ray data of HCG 62 by Ponman & Bertram (1993) and NGC 2300 by Mulchaey et al. (1993) yield different values of $a_p$ and $L_x$ than PBE (see Sect. 2).